\documentclass[pre,twocolumn,showpacs,superscriptaddress,floatfix]{revtex4}

\usepackage{graphicx}
\usepackage{latexsym} 
\usepackage{epsfig}
\usepackage{amsmath} 
\usepackage{amssymb} 
\usepackage{amsfonts}
\usepackage{hyperref}
\usepackage{bm}

\newcommand*{\pdfdxk}[3]{\frac{\partial^{#3} #1}{\partial #2^{#3}}}
\newcommand*{\pdfdx}[2]{\frac{\partial #1}{\partial #2}}
\newcommand*{\pdl}[0]{\phi_{\rm DL}}
\newcommand*{\tm}[1]{\mbox{$#1$}}

\begin{document}

\bibliographystyle{apsrev}

\title{Breathing Current Domains in Globally Coupled Electrochemical Systems:\\ A Comparison
  with a Semiconductor Model}

\date{\today}

\author{F.~Plenge}
\affiliation{Fritz-Haber-Institut der Max-Planck-Gesellschaft, Faradayweg 4--6,  D-14195 Berlin, Germany}
\author{P.~Rodin}
\affiliation{Institut f\"ur Theoretische Physik, Technische
  Universit\"at Berlin, D-10623 Berlin, Germany}
\affiliation{Ioffe Physicotechnical Institute of Russian Academy of Sciences,
Politechnicheskaya 26, 194021, St. Petersburg, Russia}
\author{E.~Sch\"oll}
\affiliation{Institut f\"ur Theoretische Physik, Technische
  Universit\"at Berlin, D-10623 Berlin, Germany}
\author{K.~Krischer}
\email{krischer@fhi-berlin.mpg.de}
\homepage{http://www.fhi-berlin.mpg.de/pc/spatdyn}
\affiliation{Fritz-Haber-Institut der Max-Planck-Gesellschaft, Faradayweg 4--6,  D-14195 Berlin, Germany}

\begin{abstract}
Spatio-temporal bifurcations and complex dynamics in globally coupled intrinsically
bistable electrochemical systems with an
S-shaped current-voltage characteristic under galvanostatic control
are studied
theoretically on a one-dimensional domain. The results are compared with the
dynamics and the bifurcation
scenarios occurring in a closely related model which describes pattern
formation in semiconductors. Under galvanostatic control both systems are unstable with respect to
the formation of stationary large amplitude current domains. The current
domains as well as the homogeneous steady state exhibit oscillatory
instabilities for slow dynamics of the potential drop across the
double layer, or across the semiconductor device, respectively.
The interplay of the
different instabilities leads to 
complex spatio-temporal behavior. We find
breathing current domains and chaotic spatio-temporal dynamics
in the electrochemical system. Comparing these findings with the results
obtained earlier for the semiconductor system, we outline
bifurcation scenarios leading to complex dynamics in globally
coupled bistable systems with subcritical spatial bifurcations.
\end{abstract}



\maketitle

\section{Introduction}
The focus of  research in nonlinear dynamics has evolved from temporal
instabilities over simple spatial patterns to complex spatio-temporal
behavior and the control or synchronization of such dynamics. Complex
spatio-temporal behavior in reaction-diffusion equations, which is in
a wider sense the class of equations dealt with also in
electrochemistry, might be found when instabilities breaking time and
space symmetries interact. A generic case is the interaction of 
Turing \cite{Turing:52} and Hopf bifurcation in a two component activator-inhibitor
system in which the involved species diffuse. Complex
spatio-temporal dynamics has been found near this codimension-two
point theoretically
\cite{Dewel.ea:95,DeWit.ea:96,Meixner.ea:97*2}
as well as experimentally \cite{Boissonade.ea:95,Valette.ea:94,HEIDEMANN.ea:93}. 

In electrochemical
systems that can be described by a two component model one variable
typically is of electrical nature and the associated transport
mechanism is migration rather than diffusion
\cite{Krischer:99,Krischer.ea:01}. The decisive variable 
for the dynamics of the electric circuit is the double layer potential
$\pdl$, measuring the voltage drop across the interface between
the working electrode and the electrolyte solution \cite{Koper:96}. Local perturbations in the
double layer potential are mediated through the electric field in the
electrolyte. Thus, spatial inhomogeneities in the double layer
potential are felt not only by its nearest neighbors, but by a whole
range of neighboring sites which makes the coupling nonlocal
\cite{Mazouz.ea:97*1}. The degree of
non-locality depends on the geometry of the electrochemical cell,
most importantly on the positions of the working (WE), counter (CE) and reference
electrode (RE) with respect to each other \cite{Mazouz.ea:97}. Furthermore it can be shown that the
galvanostatic operation mode (constant current control) introduces an additional global coupling
into the system \cite{Mazouz.ea:98}.

The role of the
second variable in two component activator-inhibitor systems in
electrochemistry is played, e.\ g., by the chemical concentration of the
reacting species  in the
double layer or by the
density of adsorbed molecules on the WE\@.

Over the last decade global coupling has been an active area of
research.
Global coupling is present in systems that are subject to
external control, e.g. via an electric circuit (such as in electrochemical
\cite{Flatgen.ea:95,Grauel.ea:98,Krischer.ea:00,Mazouz.ea:97,Mazouz.ea:98,%
Strasser.ea:00,Christoph.ea:99,Christoph.ea:99*1,Kiss.ea:99},
semiconductor
\cite{Volkov.ea:69,Bass.ea:70,Alekseev.ea:98,Meixner.ea:00*1,%
Meixner.ea:98*1,Bose.ea:00,Schimansky-Geier.ea:91,Scholl:01}
and gas discharge \cite{Willebrand.ea:92}
systems) or via the electric control of the temperature
in catalytic reactors \cite{Barelko.ea:77,Zhukov.ea:82,Graham.ea:93*1,%
Middya.ea:93,Middya.ea:94,Annamalai.ea:99}. But global coupling may also
be due to transport processes that 
happen on time scales much faster than all other relevant time scales
in the system, e.\ g.\ fast mixing in the gas phase
\cite{Mertens.ea:93,Mertens.ea:94*1,Falcke.ea:94,%
Rose.ea:96,Veser.ea:93,Falcke.ea:97}. A variety of other
systems are described by dynamics that include 
global coupling, e.g., ferromagnetic \cite{Elmer:90}, biological \cite{Seung.ea:95},
and chemical systems in which the global coupling can be light induced
\cite{Schebesch.ea:96,Epstein.ea:00}. Abstract theoretical models
are discussed, e.\ g., in \cite{Battogtokh.ea:97,Lima.ea:98,Hempel.ea:98}.

Results regarding electrochemical systems with global coupling have
been reported
for systems with an N-shaped current-voltage characteristic
(termed N-NDR systems: \underline{N}-shaped \underline{n}egative
\underline{d}ifferential \underline{r}esistance) for different types of
global coupling. In these systems the double layer potential acts as
an activator and global coupling introduced by the
galvanostatic control mode was shown to accelerate front motion
\cite{Flatgen.ea:95,Mazouz.ea:97*1} thus 
having a synchronizing effect on the spatial dynamics. Desynchronizing global coupling of the
activator was shown to stabilize potential fronts, leading to two stationary
potential domains \cite{Grauel.ea:98,Grauel.ea:01}. Also the formation of pulses and standing waves was
observed \cite{Christoph.ea:99,Strasser.ea:00,Otterstedt.ea:96}.

In electrochemical systems with an S-shaped current-voltage
characteristic (S-NDR) the roles of activator and inhibitor are
reversed, leading to a global coupling of the  inhibitor under
galvanostatic conditions. This leads
to the opposite effects opposed to N-NDR systems, i.e. current domains that are stabilized by
the constant current constraint \cite{Krischer.ea:00}.
Similar results on accelerated and decelerated fronts in globally coupled
semiconductors with S- or Z-shaped current-voltage characteristics
have also been obtained \cite{Meixner.ea:00*1}.

In the present paper we focus on the latter case of global coupling
of the inhibitor in an electrochemical system with an S-shaped
current-voltage curve. 
Furthermore, we consider systems with high electrolyte
conductivity. In such systems the migration coupling is so efficient
that any spatial variation in $\pdl$ can be neglected, which results in
 an additional global
coupling \cite{Krischer.ea:00}. The set of equations to
be investigated is thus of the general form:
\begin{eqnarray}
  \tau_{ \pdl} \pdfdx{\pdl}{t}
  &=&
  g(\phi_{\rm  DL},\langle\theta \rangle_G)  
  \label{eq:algphi}
  \\
  \tau_{\theta} \pdfdx{\theta}{t}
  &=&
  f(\phi_{\rm  DL},\theta)+D \Delta \theta,
  \label{eq:algtheta}
\end{eqnarray}
where $\theta$ stands for the activator variable, whose dynamics
comprises an
autocatalytic chemical step. The angular brackets denote the spatial average
over the spatial domain $G$. $f$ is autocatalytic in $\theta$; $g$
exhibits a monotonic characteristic as a function of $\pdl$ and
$\theta$. $\tau_{ \pdl(\theta)}$ denote the characteristic times for
changes in $\pdl$ and $\theta$, respectively.

A formally very similar set of equations describes the dynamics
in bistable semiconductor systems operated via an external load resistance
\cite{Volkov.ea:69,Bass.ea:70,Wacker.ea:94*1,Scholl:01}.
The formation and dynamics of current density patterns in bistable
semiconductors was extensively studied~\cite{Bose.ea:94,%
Wacker.ea:94,Wacker.ea:95,Alekseev.ea:98,Bose.ea:00}.
In this respective class of semiconductor systems the current-voltage
characteristic also resembles the shape of an S, which points to the
fact that the roles of the dynamic variables are very similar
to the electrochemical model:  The voltage drop $u$
across a semiconductor device acts effectively
as inhibitor, and it is subject to a global constraint
imposed by the external electric circuit. The role of the activator
variable might be played by different physical quantities, such as the
electron temperature \cite{Volkov.ea:69}, the concentration of excess
carriers \cite{Scholl:87}, the charge
density in resonant tunneling structures
\cite{Wacker.ea:95,Glavin.ea:97,Melnikov.ea:98}
(note that for bistable resonant tunneling structures the current-voltage
characteristic is Z-shaped resulting in an {\em activatory}, not
inhibitory effect of the global constraint), the voltage drop
across pn-junctions in thyristors
\cite{Gorbatyuk.ea:92,Meixner.ea:98*1} or the interface charge density in a
heterostructure hot electron diode (HHED) \cite{Wacker.ea:94}. The dynamic
equations are of the
same structural form as eq.\ (\ref{eq:algphi}),(\ref{eq:algtheta});
only the local nonlinear functions $f$ and $g$ differ from the
electrochemical model.

For the current density dynamics in a class of models
originally derived for the HHED in one or two spatial dimensions under
galvanostatic (current-controlled) conditions,
interesting complex spatio-temporal patterns termed ``spiking'' and
``breathing'' current filaments were found
\cite{Wacker.ea:94*1,Bose.ea:94}. Recently, a sufficient condition for
the onset of such complex spatio-temporal dynamics
was given \cite{Bose.ea:00}. 

Realizing the obvious similarities, we show in this paper that the
methods (e.\ g.\ for analyzing the dynamics) developed for the
semiconductor system can be applied to gain 
insight into the interaction of different instabilities in the
electrochemical system. Results regarding the possibility of the
occurrence of complex spatio-temporal behavior and the
mechanisms that lead to such behavior  are given. It is
emphasized whether the different dynamical regimes depend upon the
general structural form of the equations, especially regarding the influence
of global
coupling, or if they are due to special properties of the underlying
physical or chemical system, and thus the local dynamics.
Hence a comparison of electrochemical and semiconductor systems
gives considerable insight into generic complex dynamics of globally
coupled bistable systems.

The paper is organized as follows. In the next section we introduce
the electrochemical model, discuss its important parameters and the
mechanisms leading to global coupling in the model. In the following
section we characterize the dynamics of the model by linear stability
analysis along the lines developed for the semiconductor model and by
numerical simulations. In the discussion we compare the important
features of the two models. The mechanism leading to complex
spatio-temporal behavior in both models is different and this
difference is explored in this section in some depth. We summarize our
results in the last section and give a short outlook to applications
in terms of experimental verifications and transfer of
the electrochemical results to the semiconductor model.

\section{Model}
The central variable in electrochemical pattern formation is the
double layer potential $\pdl$, the potential drop across the interface
between the
WE and electrolyte solution. The dynamic evolution equation for
$\pdl$ can be deduced from the local charge balance at the
electrode/electrolyte interface.

\begin{figure}[!tbp]
  \begin{center}
    \epsfig{file=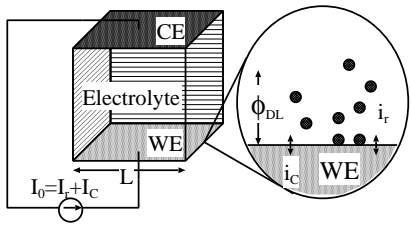,width=.3\textwidth}
    \caption{Schematic setup of the electrochemical system. A
      constant current $I_0$ is applied in a electrochemical cell
      consisting of a rectangular working electrode (WE), electrolyte,
      and a rectangular counter electrode (CE).
      WE and CE form top and bottom of the box-like cell with
      otherwise insulating walls.
      $\pdl$ is the voltage drop 
      across the interface. $i_{\rm
        r}$ and  $i_{\rm
        C}$ symbolize reaction current density and capacitive current density, respectively.}
    \label{fig:0}
  \end{center}
\end{figure}

To make things as transparent as possible and to facilitate later
comparison with the semiconductor model, we employ the
idealized geometry shown in fig. \ref{fig:0}. 
WE and CE are equally sized
rectangular plates positioned  parallel to each other in a box-like
cell with otherwise insulating walls. 
This geometry imposes 
 no-flux boundary conditions for $\pdl$ and $\theta$; there will be no
spatial inhomogeneities of the electric field at the interface imposed
by this geometry. 

For very high electrolyte conductivities $\sigma$, spatial
inhomogeneities in the double layer  potential are damped out very
fast via the efficient coupling through migration currents. It follows
that spatial variations of $\pdl$ can be neglected. This effectively
introduces a global coupling in the system, since local perturbations in $\pdl$
are felt instantaneously in the whole double layer.

In the
following we additionally assume current controlled conditions.
Galvanostatic
control is known to introduce an additional global coupling into
the system \cite{Mazouz.ea:97,Krischer.ea:00}. Assuming a specific double layer
capacitance $C$, the  dynamic equation for the double layer potential reads
\begin{equation}
  \label{eq:pdldim}
  C \pdfdx{\pdl}{t} = -i_{\rm r}(\pdl,\langle \theta \rangle) +i_0,
\end{equation}
where $i_{\rm r}(\pdl,\theta)$ is the reaction current density and $i_0$
denotes the imposed current density. 
The activator variable $\theta$ describes the evolution of the
coverage of the WE by an adsorbate or the concentration of a
chemical species in the reaction plane. Its dynamics will be modeled
by an equation of the form (\ref{eq:algtheta}), where we restrict our system to
one spatial dimension (1d) since the qualitative behavior should
also be captured on 1d domains. 1d domains also 
resemble the situation of a very large aspect ratio of the
rectangular domain, where one spatial dimension is too small to allow
for spatial instabilities and can thus be eliminated.

We use the following model functions for the local
dynamics of the activator and the reaction current density $i_{\rm r}$
\begin{align}
  \label{eq:functheta}
  i_{\rm r}(\pdl,\theta)
  & = 
  (1-\theta) e^{\pdl}\\
  \label{eq:funcir}
  f(\pdl,\theta)
  &=
  (1-\theta) e^{-\nu \pdl^2 - g \theta} - p \theta
  e^{\nu \pdl^2 + g \theta}
\end{align}
originally derived to describe pattern formation observed in a
reaction, in which a reaction inhibiting adsorbate undergoes a first order phase
transition due to lateral interactions of the adsorbate molecules
\cite{Mazouz.ea:00,Li.ea:01}. The transformations leading to
dimensionless units differ from the
ones given in \cite{Mazouz.ea:00}; the derivation is
given in 
Appendix \ref{app1}. Note the non-polynomial nature of the
function $f$. 

The dimensionless set of equations is thus
\begin{align}
  \label{eq:nondimphi}
  \pdfdx{\pdl}{t}
  &=
  \gamma(i_0-(1-\langle\theta \rangle) e^{\pdl})
  \\
  \pdfdx{\theta}{t}
  &=
  \mu \left((1-\theta) e^{-w(\theta,\pdl)}
  - p \theta
  e^{w(\theta,\pdl)} \right) + \pdfdxk{\theta}{x}{2}
  \label{eq:nondimtheta}
  \intertext{with}
  w(\theta,\pdl)
  &=
  \nu \pdl^2 + g \theta  \nonumber,
\end{align}
subject to the boundary conditions
\begin{equation*}
  \left.\pdfdx{\theta}{x}\right|_{x=0,\pi}=0.
\end{equation*}
We normalize space to the interval $[0,\pi]$ for computational
convenience and thus 
\begin{equation*}
  \langle \theta \rangle= \frac{1}{\pi}
  \int_0^{\pi} \theta(x) \, {\rm d} x.
\end{equation*}
This leads to the proportionality of the parameters \mbox{$\mu,
\gamma \propto L^2$}; $\mu$ and $\gamma$ still can be changed independently since
also other physical quantities enter these parameters (cf.\ Appendix \ref{app1}).

\begin{figure}[!tbp]
  \begin{center}
    \epsfig{file=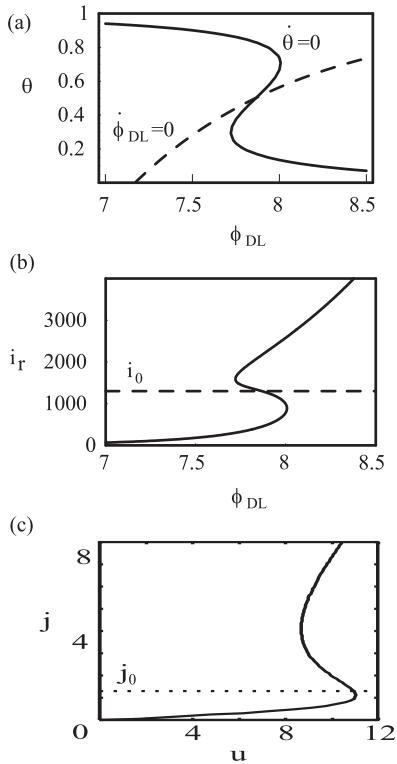,width=.29\textwidth}
    \caption{(a) Nullclines of the model
      (\ref{eq:nondimphi}),(\ref{eq:nondimtheta}) for an imposed
      current density 
      in the autocatalytic regime (solid line \tm{\dot{\theta}=0},
      dashed line \tm{\dot{\pdl}=0}, $i_0=1300$, for the other
      parameters see Appendix \ref{app1}). (b),(c) S-shaped current-voltage
      curve together with the load line $i_0$  ($j_0$) for the
      electrochemical (eq.\ (\ref{eq:nondimphi}),(\ref{eq:nondimtheta})) and the
      semiconductor (eq.\ (\ref{eq:sema}),(\ref{eq:semu}))
      system, respectively.}
    \label{fig:1}
  \end{center}
\end{figure}

In fig. (\ref{fig:1}a)
 the nullclines of the system are depicted for a
current density $i_0$ that is set in the range of the negative differential
resistance in the current-voltage characteristic (see fig. \ref{fig:1}b). The S-shaped
current-voltage-characteristic is depicted together with the load
line $i=i_0$ in fig.\ \ref{fig:1}b. This physically more intuitive
\mbox{($i$-$\pdl$)}-plane representation will  be used in the
following. 

The
parameters $\nu$, $p$ and $g$ are fixed throughout this paper at the
values $\nu=0.025$, $p=0.5$ and $g=-2.4$ (cf.\ Appendix \ref{app1}).
The dynamics is determined by the model parameters $\mu$, essentially
proportional to $L^2$, the
relaxation time ratio of activator and inhibitor $\gamma/\mu$
(independent of $L$), and the
general excitation level controlled by the imposed current density $i_0$. The relaxation
time ratio can be accessed easily via the concentrations of the
reacting and adsorbing species; $i_0$ is set by the galvanostatic
control unit. 

The numerical results discussed in section \ref{homlistatdo} were obtained
using pseudo spectral decomposition in space \cite{canuto:93} employing 15
spatial cosine-modes (the results do not change when a larger number
of modes is chosen). For the integration in time the routine lsode \cite{Hindmarsh:80}
and for continuation of steady states and limit cycles the package AUTO \cite{Doedel.ea:86}
was used. 

\section{Stability Analysis and Simulations}

\subsection{Homogeneous Steady State}
\label{Homogeneous Steady State}
In this section we consider the spatially uniform fixed points of the
system (\ref{eq:nondimphi}),(\ref{eq:nondimtheta}) and their
bifurcations. The uniform steady state $(\pdl^{\rm ss},\theta^{\rm
  ss})$ is given by \tm{i_r(\pdl^{\rm ss},\theta^{\rm ss})=i_0},
\tm{f(\pdl^{\rm ss},\theta^{\rm ss})=0} 
 and corresponds to the homogeneous
S-shaped current-voltage characteristic
(fig.\ \ref{fig:1}b). Perturbing the steady state  with a perturbation
$(\delta \pdl e^{\lambda t},\delta \theta \cos (nx) e^{\lambda t} )$
(consistent with the boundary conditions), the temporal evolution of
the perturbation is given by the eigenvalues of the Jacobian matrix $J$
\begin{equation*}
  \lambda_{1,2}=\frac{{\rm tr} J}{2} \pm \sqrt{\frac{({\rm tr} J)^2}{4}-\det J}
\end{equation*}
and stability $({\rm Re}\lambda < 0)$ implies that $(\det J>0
\quad \wedge \quad {\rm tr} J<0)$. The Jacobian reads
\begin{equation*}
  J=
  \begin{pmatrix}
    - \gamma \sigma_{\rm r} & - \gamma i_{\rm r_{\theta}}\\
    \mu f_{\pdl} & \mu f_{\theta}-n^2
  \end{pmatrix}.
\end{equation*}
Subscripts denote partial derivatives with respect to the
subscripted variable and evaluation at the steady state
(e.\ g.\ \tm{f_{\theta}=\left. \pdfdx{f}{\theta} \right|_{(\pdl^{\rm ss},\theta^{\rm
    ss})}}). For brevity we denote \tm{\sigma_{\rm r} := \left. \pdfdx{i_{\rm r}}{\pdl}
  \right|_{(\pdl^{\rm ss},\theta^{\rm ss})}}.

The stability of the fixed point with respect to homogeneous
fluctuations \mbox{$(n=0)$} can be determined by inspecting 
\begin{eqnarray*}
  \label{eq:detj}
  \det J 
  &=& 
  - \gamma \mu f_{\theta} ( \sigma_{\rm r}- \frac{f_{\pdl}}{f_{\theta}}
  i_{\rm r_{\theta}})
  \\
  &=& - \gamma \mu  f_{\theta} \left( \sigma_{\rm r} + i_{\rm r_{\theta}} \frac{{\rm d}
    \theta^{\rm ss}(\pdl)}{{\rm d} \pdl} \right)\\
  &=& - \gamma \mu  f_{\theta} \frac{{\rm d}
    i_r(\theta^{\rm ss}(\pdl),\pdl)}{{\rm d} \pdl}\\
\end{eqnarray*}
and
\begin{equation*}
  \label{eq:trj}
  {\rm tr} J = \mu f_{\theta}-\gamma \sigma_{\rm r}.
\end{equation*}
Obviously \mbox{$\det J > 0$} in general since
\mbox{$\mu,\gamma>0$} and \mbox{$f_{\theta} \frac{{\rm d}
    i_r(\theta^{\rm ss}(\pdl),\pdl)}{{\rm d} \pdl}<0$}, which follows
from the fact that the branch of negative differential resistance (\mbox{$\frac{{\rm d}
    i_r(\theta^{\rm ss}(\pdl),\pdl)}{{\rm d} \pdl}<0$}) is caused
solely by the activator variable $\theta$, equivalent to saying that
\mbox{$\sigma_{\rm r}>0$} in general.

\begin{figure}[!tbp]
  \begin{center}
    \epsfig{file=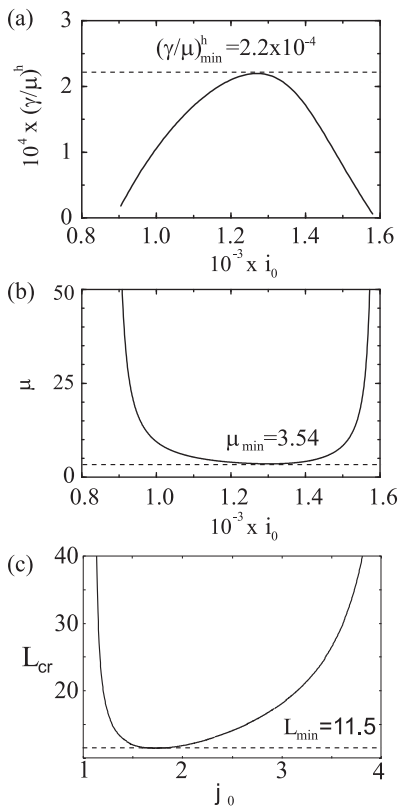,width=.29\textwidth}
    \caption{(a) Location of the Hopf-bifurcation  of the
      homogeneous steady state in the ($\gamma/\mu$-$i_0$)-parameter-plane
      for the electrochemical model (\ref{eq:nondimphi}),(\ref{eq:nondimtheta}). For
      $\gamma/\mu>2.2\cdot10^{-4}$ the system is stable with respect
      to homogeneous fluctuations. (b) Threshold for the spatial instability of the uniform
      steady state in the ($\mu$-$i_0$)-plane. For system sizes
      smaller than $\mu_{\rm min}=3.54$ the system is stable with
      respect to spatial fluctuations. (c) Critical system size
      $L_{\rm cr}$, of
      the spatial instability for the semiconductor model
      (eq.\ (\ref{eq:sema}),(\ref{eq:semu}))
      as a function of the imposed current density $j_0$.}
    \label{fig:2}
  \end{center}
\end{figure}

However, \mbox{tr $J$} might change sign on the NDR-branch since
\mbox{$f_{\theta}>0$} and \mbox{$\sigma_{\rm r}>0$}, which leads to an
oscillatory instability (Hopf bifurcation) of the homogeneous steady
state (denoted by a superscript ``h'', cf.\ table \ref{tab:2}) at
\begin{table}
\caption{Abbreviations for bifurcation points}
\label{tab:2}
\begin{ruledtabular}
\begin{tabular}{l p{0.8\linewidth}}
h&Hopf bifurcation of the homogeneous steady state\\
d&domain bifurcation of the homogeneous steady state\\
sn-d&saddle-node bifurcation of domains\\
hd&Hopf bifurcation of the domain state\\
snp&saddle-node bifurcation of breathing domains,
i.\ e.\  periodic orbits\\
DH&domain-Hopf codimension-two point (d and h)\\
TB&Takens-Bogdanov codimension-two point (sn-d and hd)\\
DHD&degenerate Hopf bifurcation of domains (snp and hd)
\end{tabular}
\end{ruledtabular}
\end{table}
\begin{equation}
  \label{eq:hopf}
  \left(\frac{\gamma}{\mu}\right)^{\rm h} = \frac{f_{\theta}}{\sigma_{\rm r}}.
\end{equation}
Thus for \mbox{$\gamma/\mu<\left(\gamma/\mu\right)^{\rm
    h}$} (low concentration of the reacting species or high
concentration of the adsorbate) the homogeneous steady state is
unstable in a certain $i_0$-interval, since
\mbox{$f_{\theta} \sigma_{\rm r}^{-1}$} depends on the imposed
current density via the steady state condition. When plotting the critical value
\mbox{$\left(\gamma/\mu\right)^{\rm h}$} as 
a function of the imposed current density fig.\ \ref{fig:2}a is obtained.

For
\tm{\gamma/\mu > (\gamma/\mu)^{\rm h}_{\rm max} = 2.2\cdot10^{-4}} there are no oscillatory
solutions for any $i_0$ and for \tm{\gamma/\mu \ll  (\gamma/\mu)^{\rm h}_{\rm max}}
the oscillatory instability takes place close to the turning points of
the current-voltage characteristic at \mbox{$i_0=889$} and \mbox{$i_0=1587$}.

To determine the stability with respect to spatially inhomogeneous
fluctuations, it is sufficient to consider the activator variable
$\theta$, since sinusoidal perturbations do not affect the average value
of $\theta$ and thus the double
layer dynamics. Therefore the steady state becomes unstable with
respect to the \mbox{$n^{\rm th}$-mode} for
\begin{equation*}
  \mu>\frac{n^2}{f_{\theta}},
\end{equation*}
and the first mode to become unstable is always the mode with
wavenumber one \cite{Mikhailov:90}. The
wavelength of the first unstable mode depends on the system size and
is equal to $2L$ for Neumann boundary conditions. In the following we term this
instability \emph{domain bifurcation}. The critical
parameter value is thus:
\begin{equation}
  \label{eq:mucrit}
  \mu^{\rm d} = f_{\theta}^{-1}.
\end{equation}
This critical value is depicted in fig.\ \ref{fig:2}b as a function of
$i_0$. For systems sizes \mbox{$\mu < \mu_{\rm min}=3.54$} the spatial instability is
suppressed; this defines a natural length scale for the system. For
system sizes much larger than this natural length scale the spatial
instabilities occur once again close to the turning points of the
current-voltage characteristic.

The spatial and oscillatory instabilities may coincide in a
codimension-two point (Domain-Hopf bifurcation, ``DH'', cf.\ table \ref{tab:2}) if 
\begin{equation}
  \label{eq:ct}
  \gamma^{\rm DH}=\sigma_{\rm r}^{-1}.
\end{equation}
The respective imposed current density value $i_0^{\rm DH}(\mu)$
is defined as the solution of (\ref{eq:mucrit}) with respect to $i_0$.

\subsection{Homogeneous Limit Cycle and Stationary Domains}
\label{homlistatdo}
In this section we complete the picture of the different basic attractors of
the model by including limit cycles and stationary current domains
into our stability analysis. Analytical methods fail in most cases since the
involved bifurcations are either subcritical and thus do not allow for
an amplitude equation analysis and/or the considered system sizes
are intermediate, which excludes methods like singular
perturbation theory \cite{Kerner.ea:94} to describe  domain interface dynamics.

\begin{figure}[!tbp]
  \begin{center}
    \epsfig{file=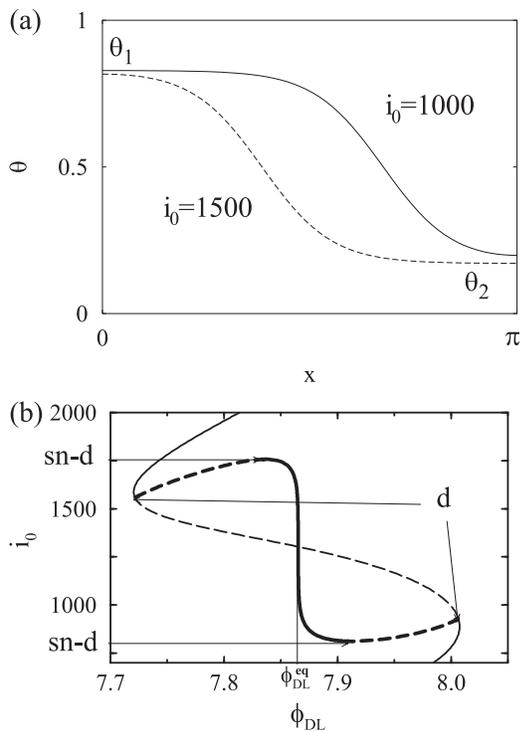,width=.38\textwidth}
    \caption{(a) Stable domains for two values of the imposed current density
      $i_0$ for the electrochemical model (\tm{\mu=25},
      \tm{\gamma=10}). (b) Bifurcation diagram for 
       \tm{\mu=25} and  \tm{\gamma=10}. Shown is $\pdl$ as a
      function of the bifurcation parameter $i_0$ in the familiar
      current-potential plane. The branch of negative differential
      resistance is unstable (thin dashed line) with respect to domain
      formation. The domain branches (thick lines) bifurcate
      subcritically (d) near the
      turning points of the current-voltage characteristic. The
      stable and unstable domain branches (solid resp.\ dashed thick
      lines) are born in a saddle-node 
      bifurcation of domains (sn-d). The domain branch  can be approximated
      by an equal-areas rule, eq.\ (\ref{eq:equalareas}), in a huge $i_0$
      interval yielding an equistability potential  \tm{\pdl^{\rm eq}}.}
    \label{fig:3}
  \end{center}
\end{figure}

For common concentrations and system sizes the double layer dynamics
will be much faster than the dynamics of the activator. For these
conditions the parameters $\gamma$ and $\mu$ will be of the order
10 and 100, respectively. It follows that in most cases
oscillatory instabilities are not present in the system and the only
nontrivial mode is a stationary current domain as depicted
in fig.\ \ref{fig:3}a
 for two values of $i_0$. This current
domain is the final state of the system in the spatially unstable
regime and the mechanism leading to such a stationary domain is well
known (e.\ g. \cite{Mikhailov:90,Krischer.ea:00}): 

The activator is bistable as a function of the double layer
potential.
An overcritical local fluctuation  in a system without global coupling that is
prepared in the metastable state  would lead to the formation
of a transition front to the globally stable state. The global
constraint, however, forces the system to maintain an 
average current. The system meets this constraint by taking on an
inhomogeneous state in which 
two phases coexist. With other words, the front velocity becomes zero. The final
state of the system is described by a Maxwell type construction: the
intermediate, equistability  double layer potential $\pdl^{\rm eq}$, which is
established in the stationary structure
 is determined by the equal-areas rule \cite{Mikhailov:90,Scholl:01}
\begin{equation}
  \label{eq:equalareas}
  \int_{\theta_1}^{\theta_2} f(\pdl^{\rm eq},\theta) \, {\rm d} \theta =0.
\end{equation}
In fig.\ \ref{fig:3}b the bifurcation diagram with respect to $i_0$ is
shown for \mbox{$\mu=25$}, \mbox{$\gamma=10$}. Even though the system
size is comparable to the interface width, as can be seen in
fig.\ \ref{fig:3}a, the above construction holds for a wide
$i_0$-interval. However, since the arguments given
above apply strictly only for infinite systems, deviations
near the turning points of the current-voltage characteristic of the
domain are clearly visible.
These deviations represent a boundary effect.

States with several domains are unstable due to the winner takes all
principle \cite{Volkov.ea:69,Schimansky-Geier.ea:91}. Domains with an extremum
not located at the boundaries are unstable with respect to translation and
will be attracted by the boundary.

Fig.\ \ref{fig:3}b also shows that spatially patterned
solutions typically bifurcate subcritically from the homogeneous state
and meet the 
stable domain-branch in a saddle-node bifurcation (sn-d,
 cf.\ table \ref{tab:2}). 
The domains remain stable in the whole
$i_0$-interval in which the current-voltage characteristic of the
domain exhibits a negative
differential resistance for these parameter values. This behavior can also be rationalized
analytically \cite{Bass.ea:70,Alekseev.ea:98}.
The domain bifurcation is supercritical only in a small $\mu$-interval close
to the minimal system size $\mu_{\rm min}$.

\begin{figure}[!tbp]
  \begin{center}
    \epsfig{file=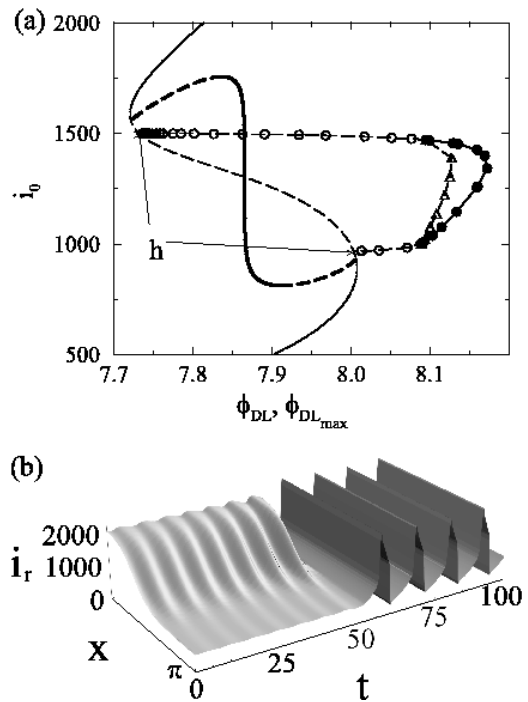,width=.38\textwidth}
    \caption{(a) Bifurcation diagram for the electrochemical model
    (\tm{\mu=25},
      \tm{\gamma=3\cdot10^{-3}}). Apart from the branches shown in
      fig.\ \ref{fig:3}b, a branch of unstable homogeneous oscillatory solutions
      (open circles) bifurcates supercritically (h)
      near the turning points of the current-voltage characteristic.
      Shown is the maximum value of $\pdl$ during one oscillation
      cycle. After stabilization through a pitchfork bifurcation, the
      stable homogeneous oscillations 
      (full circles)
      are separated from the stable domains by an unstable inhomogeneous limit
      cycle (open triangles). (b) Typical scenario of an oscillatory
      instability of a domain for lower values of $\gamma$ than in
      (a)  (\tm{\mu=25}, \tm{\gamma=1\cdot10^{-4}},
      \tm{i_0=1000}). Shown is the reaction current density 
      \tm{i_r=(1-\theta)e^{\pdl}} as a function of space and
      time. At these parameter values the oscillatory instability of
      the domain is subcritical and the system finally settles down
      to homogeneous relaxation
      oscillations (standard scenario). }
    \label{fig:4}
  \end{center}
\end{figure}

When $\mu$ is fixed at a value \tm{\mu>\mu_{\rm min}} and  the double
layer dynamics is slowed down to $\gamma$ below 
\tm{\mu (\gamma/\mu)^{\rm h}_{\rm max}}, the additional
mode of homogeneous oscillations becomes present in the system. For
\tm{\gamma \lesssim \mu (\gamma/\mu)^{\rm h}_{\rm max}}
it bifurcates  supercritically from the spatially unstable state, therefore small amplitude
oscillations will be unstable with respect to spatial
fluctuations for any $i_0$. With increasing oscillation amplitude (decreasing
$\gamma$) the oscillations become stabilized in a pitchfork
bifurcation (cf.\ fig.\ \ref{fig:4}a). 
This results in bistability
of stationary domains and an uniform limit cycle in an intermediate
$i_0$-interval. The basins of 
attraction are separated by an unstable inhomogeneous limit
cycle.

If $\gamma$ is lowered even further, the stationary current
domain will become unstable also. This can be rationalized by recalling
that the stabilization mechanism of the domains is the positive global
coupling on $\pdl$. If the delay of the double layer dynamics
becomes too large, $\pdl$ can no longer control the interface stability. We
denote the critical value of this oscillatory instability of the domain by
$\gamma^{\rm hd}(\mu,i_0)$ (cf.\ table \ref{tab:2}). Numerical
simulations show that  the  threshold 
for an 
oscillatory instability of the current domain lies typically below the
threshold for the Hopf bifurcation of the homogeneous steady state 
\begin{equation}
  \label{eq:gammacrit 1}
  \gamma^{\rm hd}(\mu,i_0)<\gamma^{\rm h}(\mu,i_0).
\end{equation}
This can be understood in the frame of the eigenmodes of the current domain
for large system sizes if we recall that in absence of global coupling
the domain state has only one positive eigenvalue that tends to zero with
increasing system size. The respective arguments are given in
\cite{Alekseev.ea:98}. The numerical investigations show that 
relation (\ref{eq:gammacrit 1}) in general holds for small and
intermediate system sizes also. 

It follows that, in general, the homogeneous relaxation oscillations
represent an attractor when the domain loses stability. The
oscillatory instability of the domain is usually subcritical; a state
close to the domain is eventually attracted by the
stable homogeneous limit cycle (see fig.\ \ref{fig:4}b). This can be regarded as the standard
scenario (i.\ e.\ it exists in a wide parameter range)  of
a domain instability in globally coupled
electrochemical systems with an S-shaped current-voltage
characteristic.
In this case no complex spatio-temporal behavior arises
in the model.

\subsection{Breathing Current Domains}

\begin{figure}[!tbp]
  \begin{center}
    \epsfig{file=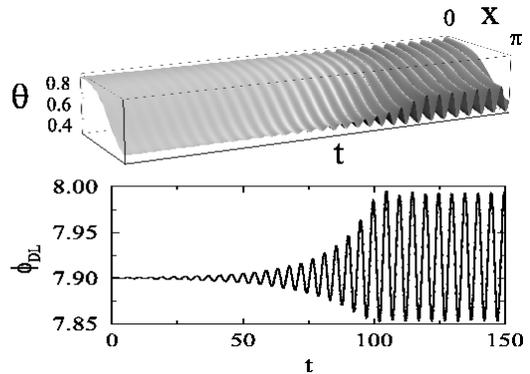,width=.38\textwidth}
    \caption{Oscillatory instability of a domain leading to stable,
      periodically breathing current domains
      for the electrochemical model (\tm{\mu=10},
      \tm{\gamma=7\cdot10^{-4}}, \tm{i_0=1000}). (In this simulation a stable domain was
      prepared, $\gamma$ was lowered to
      \tm{\gamma=7\cdot10^{-4}}, and a small random fluctuation was added.)}
    \label{fig:5}
  \end{center}
\end{figure}

\begin{figure}[!tbp]
  \begin{center}
    \epsfig{file=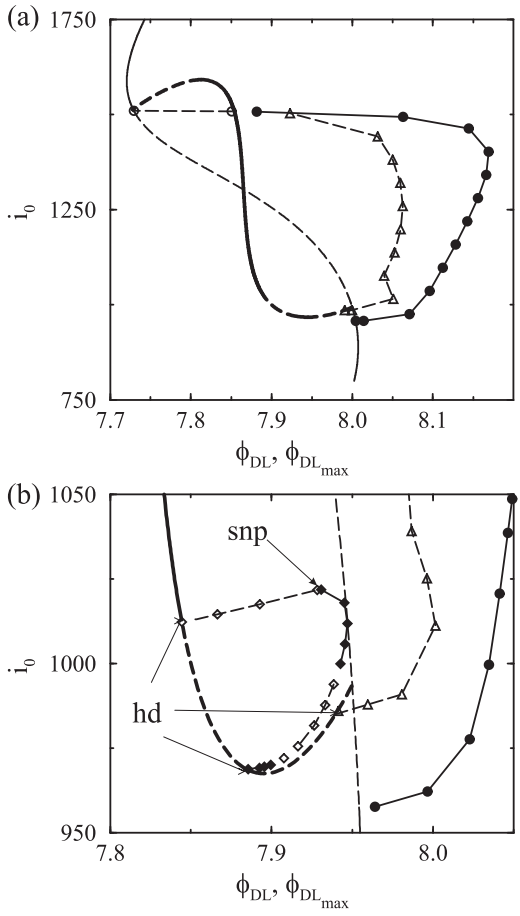,width=.38\textwidth}
    \caption{(a) Basis bifurcation diagram for stable periodic
      breathing for the electrochemical model (\tm{\mu=10},
      \tm{\gamma=7\cdot10^{-4}}). The oscillatory
      branch of the homogeneous limit cycles bifurcates supercritically
       before the spatial instability and thus
      homogeneous oscillations are stable nearly in the whole
      $i_0$-interval (full circles). The equal areas rule, eq.\
      (\ref{eq:equalareas}), fails for this 
      system size. The domain branch (thick line) is unstable in a region of
      negative differential resistance (dashed thick line) near the lower saddle
      node bifurcation of domains. Marked with open triangles is an
      unstable inhomogeneous limit cycle. It is born in a pitchfork
      bifurcation of the homogeneous limit cycle at high $i_0$ and
      terminates in the unstable domain branch. (b) Enlargement
      of the bifurcation diagram at the lower turning point. Here also
      the branches of the inhomogeneous breathing mode are shown (diamonds).
      The breathing mode bifurcates subcritically (hd) from the domain branch at higher
      $i_0$ (open diamonds) and stable breathing (full diamonds) originates in an snp. In the
      projection of the limit cycle on the double layer potential it gets close to
      the homogeneous steady state but not in real phase space (cf.\ 
      text). In the current density interval between approx. \tm{i_0=1000} and
      \tm{i_0=975} the inhomogeneous limit cycle undergoes a period
      doubling cascade leading to chaotic breathing (open diamonds).}
    \label{fig:6}
  \end{center}
\end{figure}

We would expect complex spatio-temporal behavior if the branch of
inhomogeneous limit cycle solutions that bifurcates from the
domain-branch at the point of the  oscillatory instability of the
domain becomes stabilized or bifurcates supercritically. In this case
the system would exhibit bistability 
between a stable inhomogeneous limit cycle and a stable homogeneous
one. We did indeed find such a situation in the model for
comparatively small system sizes (\mbox{$\mu \sim 10$}) and relaxation
times well below the onset of homogeneous oscillations
(\mbox{$\gamma/\mu \sim 7\cdot10^{-5}$}). 
The instability leading to
such complex spatio-temporal behavior is shown in
fig.\ \ref{fig:5}. 
In fig.\ \ref{fig:6}a 
the corresponding
bifurcation diagram for \mbox{  $\mu = 10$} and  \mbox{$\gamma =
  7\cdot10^{-4}$}  is depicted. 

\begin{figure}[!tbp]
  \begin{center}
    \epsfig{file=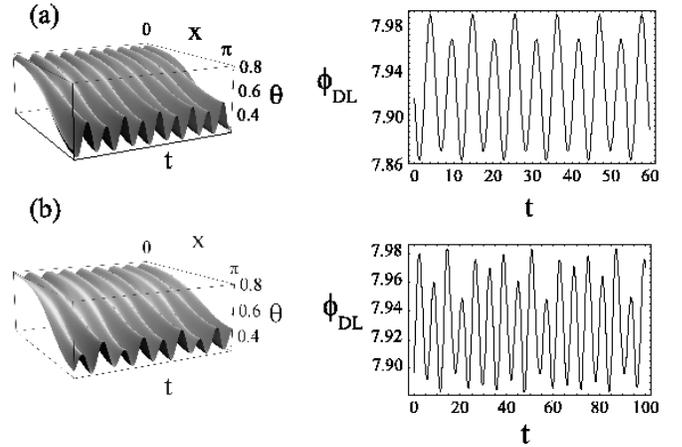,width=.48\textwidth}
    \caption{(a) Periodically breathing domains with period two at
      \tm{i_0=990} and (b)  chaotically 
      breathing domains at \tm{i_0=980} (\tm{\mu=10},
      \tm{\gamma=7\cdot10^{-4}}).}
    \label{fig:7}
  \end{center}
\end{figure}

Decreasing the imposed current density from values in the regime of
bistability between a stable domain and homogeneous oscillations, the
domain branch exhibits an oscillatory instability (hd).
The branch of solutions
that bifurcates subcritically is stabilized via a saddle-node
bifurcation of oscillatory domains, i.\ e.\  periodic orbits (snp,
cf.\ table \ref{tab:2}), which can be seen in the 
enlarged bifurcation diagram, fig.\ \ref{fig:6}b. 
The spatio-temporal behavior becomes more involved as the imposed
current density $i_0$ is decreased. The limit cycle undergoes a period
doubling cascade  leading to stable chaotic spatio-temporal
motion (fig.\ \ref{fig:7}).
Decreasing $i_0$ further, a reversed period doubling cascade occurs
which leads again to stable period one breathing domains. This branch 
then ends in a supercritical Hopf bifurcation of the domain very close
to the saddle-node bifurcation, in which stable and unstable domains
originate (sn-d). 
It is interesting to note that the
dynamic nature of the invariant set that separates the basins of
attraction of the two limit cycles is changing with increasing imposed
current density from the unstable stationary domain (saddle point) to an
unstable inhomogeneous limit cycle (see fig.\ \ref{fig:6}b for low $i_0$). 

\begin{figure}[!tbp]
  \begin{center}
    \epsfig{file=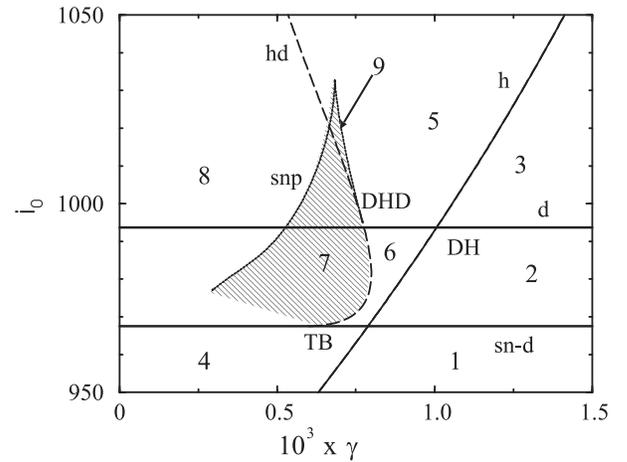,width=.44\textwidth}
    \caption{Existence region of stable periodic breathing in the
      ($i_0$-$\gamma$)-control-parameter plane (hatched region) for
      the electrochemical model (\tm{\mu=10}). 
      The main dynamic regimes, characterized by attractors, 
      are indicated by the numbers 1--9; the attractors are given in table \ref{tab:3}. 
      Shown in solid lines are the lines of the domain
      bifurcation (d) and the saddle-node bifurcation in which the
      domains originate (sn-d) (both independent of $\gamma$). The
      domain bifurcation and the
      Hopf bifurcation of the homogenous steady state (h, solid)
      intersect in a Turing-Hopf-type codimension-two point
      (DH). The dashed line shows the oscillatory instability of the
      domain (hd). Denoted by snp (solid line) is the line of the saddle node
      bifurcation of periodic orbits, i.\ e.\ breathing domains.}
    \label{fig:8}
  \end{center}
\end{figure}

The region in the ($i_0$-$\gamma$)-parameter-plane
in which such complex spatio-temporal dynamics is found is depicted in
fig.\ \ref{fig:8} for \tm{\mu=10}.
\begin{table}[!tbp]
  \begin{center}
    \caption{Dynamic regimes indicated in fig.\ \ref{fig:8}}
    \begin{ruledtabular}
      \begin{tabular}{c p{0.9\linewidth}}
        (1)&    One stable homogeneous fixed point.\\
        (2)&    Bistability between stable domain and homogeneous fixed point.\\
        (3)&    Only one stable domain.\\
        (4)&    One stable homogeneous limit cycle.\\
        (5)&    Stable or unstable homogeneous limit cycle (cf.\ Appendix B) and stable
        domain.\\
        (6)&    Stable homogeneous limit cycle and stable domain.\\
        (7)&    Stable breathing current domains (periodic or chaotic)
        and a stable homogeneous limit cycle.\\
        (8)&    The Hopf bifurcation of the domain is subcritical, thus only stable
        homogeneous  oscillations are present.\\
        (9)&    Region in which three attractors exist
        (cf.\ fig.\ \ref{fig:6} for $i_0 \cong 1010$):
        Stable domains, stable breathing domains, and stable
        homogeneous limit cycle.
      \end{tabular}
    \end{ruledtabular}
    \label{tab:3}
  \end{center}
\end{table}
The lines of the Hopf bifurcation and the domain bifurcation of the homogeneous
steady state and their intersection point (DH) are shown. The main
regions that were discussed above (and in part also exist for
different values of $\mu$) are indicated. Note the existence of three
codimension-two points: The point in which domain and Hopf bifurcation
coincide (DH) was discussed in sec.\ \ref{Homogeneous Steady State}. At
the DH the system has a pair of purely imaginary eigenvalues and a real
eigenvalue equal to zero \cite{Guckenheimer.ea:83}. 
Unfoldings of the DH have a further fine structure as discussed in
Appendix \ref{app2}; it is  not shown in fig.\ \ref{fig:8} for clarity.
Denoted by TB is the point where saddle-node and Hopf-bifurcation
meet (Takens-Bogdanov point) \cite{Guckenheimer.ea:83}. Note that in our case 
both bifurcations involve inhomogeneous steady states (i.\ e.\ domains)
rather than homogeneous solutions. Left of the TB two
saddle fixed points with one and three unstable directions,
respectively, originate from the saddle-node bifurcation, right of it
a 
saddle fixed point and a stable node. Again the fine structure, most remarkably a
homoclinic bifurcation that should be present in the vicinity of the
TB, is omitted. The third codimension-two point is a
degenerate Hopf bifurcation of domains (DHD), in which the
saddle-node bifurcation of periodic orbits (snp) coincides with the
Hopf-bifurcation of the domain (hd).

We omitted in the bifurcation diagram (fig.\ \ref{fig:8}) some of the
branches mentioned above. Furthermore there are indications of the
presence of additional bifurcations that determine the exact location
of the lower boundary of the regime of complex behavior. Its detailed
study is beyond the scope of this paper.

\section{Comparison and Discussion}
In this section we compare the different dynamic instabilities and
regimes described in the previous section with results obtained
earlier for the semiconductor model. The semiconductor model used has
the (nondimensionalized) form
\begin{align}
  \label{eq:sema}
  \dot{a}&=\frac{u-a}{(u-a)^2+1}-0.05 a + \pdfdxk{a}{x}{2}\\
  \label{eq:semu}
  \dot{u}&=\alpha(j_0-u+\langle a \rangle_L),
\end{align}
where $u$ denotes the potential drop across the semiconductor device
(corresponding to $\pdl$) and $a$ describes the interface charge
density in the HHED (corresponding to $\theta$). The system length is $L$ and thus \mbox{$\langle a
  \rangle_L=L^{-1}\int_0^L a \, {\rm d} x$}. The current-voltage
characteristic of the HHED is given by
\tm{j=u-a}. It also has the shape of an `S' (fig.\ \ref{fig:1}c). If space is rescaled to the
interval $[0,\pi]$, the model
exhibits the same structural dependence on three parameters as
eqs.\ (\ref{eq:nondimphi}),(\ref{eq:nondimtheta}) 
\begin{align}
  \dot{a}&=\mu^{\rm (s)} \left( \frac{u-a}{(u-a)^2+1}-0.05 a \right) + \pdfdxk{a}{x}{2}\\
  \dot{u}&=\gamma^{\rm (s)}(j_0-u+\langle a \rangle_{\pi}),
\end{align}
with \tm{\mu^{\rm (s)}=\left(\frac{L}{\pi}\right)^2} and \tm{\gamma^{\rm (s)}
  = \left(\frac{L}{\pi}\right)^2 \alpha}. These parameters can be
interpreted in the same way as in the
electrochemical model.

\begin{figure}[!tbp]
  \begin{center}
    \epsfig{file=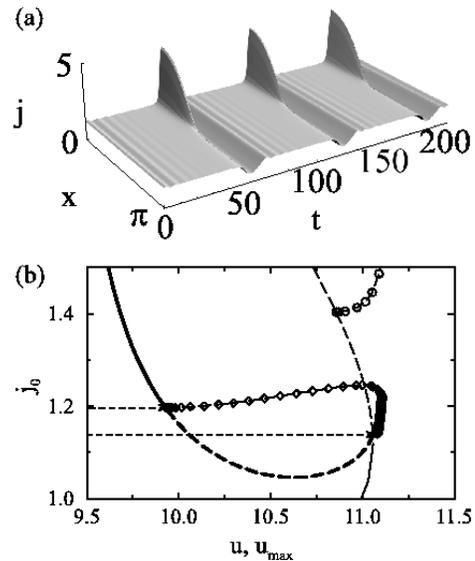,width=.34\textwidth}
    \caption{(a) Spiking current filament in 1d (\tm{L=40},
      \tm{\alpha=0.035}, \tm{j_0=1.2}). (b) Bifurcation diagram for the
      semiconductor system for complex spatio-temporal dynamics (\tm{L=40},
      \tm{\alpha=0.06}). Shown is the potential drop across the
      semiconductor $u$ resp.\ the maximum $u$ during one oscillation
      versus the imposed current density
      $j_0$ at the lower turning point of the current-voltage
      characteristic. In the current-interval shown by the dashed
      lines no
      trivial state of the system is stable. The lower boundary is
      the spatial instability of the homogeneous steady state (thin lines) and
      the upper one is the oscillatory instability of the
      filament (thick lines). Homogeneous oscillations are not present in this
      current density interval, they bifurcate at higher current density values
      (open circles in the upper right corner).
      The resulting inhomogeneous oscillations (diamonds) that
      bifurcate subcritically from the stable domain branch are
      born by a saddle-node bifurcation of periodic orbits.}
    \label{fig:9}
  \end{center}
\end{figure}

The two models possess equivalent basic modes: The branch of negative differential conductivity
is unstable with respect to spatial perturbations for sufficiently large
system sizes \tm{L>L_{\rm min}} (cf.\ fig.\ \ref{fig:2}c) and to homogeneous oscillations for
sufficiently slow dynamics of the voltage drop $u$ (small $\alpha$). However the temporal
instability of the filament may lead
to qualitatively different spatio-temporal dynamics. Apart
from the breathing mode that the semiconductor models also exhibit,
\cite{scholl.ea:90,Wacker.ea:94,Bose.ea:00,Scholl:01}
the system displays a complex spatio-temporal mode termed {\em spiking}
(see fig.\ \ref{fig:9}a) \cite{Wacker.ea:94*1,Bose.ea:94}.

This mode evolves because the
spatially inhomogeneous limit cycle that constitutes breathing comes
eventually, with decreasing $\gamma$, very close to the homogeneous fixed point.
This points to a structurally different dynamic regime as compared to
the electrochemical system and facilitates the formulation of a
sufficient condition for the occurrence of complex spatio-temporal
dynamics \cite{Bose.ea:00}.
In the following this will be explained in some detail.

Consider the bifurcation diagram of the semiconductor model for
parameter values at which complex spatio-temporal dynamics is found
(fig.\ \ref{fig:9}b). Let us denote by \tm{j_0^{\rm d}(\mu)}, \tm{j_0^{\rm
  h}(\mu,\gamma)} and \tm{j_0^{\rm hd}(\mu,\gamma)} the parameter
values at which the spatial instability of the homogeneous steady
state, the oscillatory instability of the homogeneous steady state and
the oscillatory instability of the filament, respectively, occur.
For an interval of imposed current densities $j_0$ no trivial state is
stable, since, in contrast to the electrochemical model, homogeneous
oscillations are not present in the system for imposed current density values
within this interval, however the filament is already oscillatorily unstable
(\tm{\gamma^{\rm hd}(\mu,i_0)>\gamma^{\rm h}(\mu,i_0)}). Thus a
sufficient condition for complex dynamics is 
\begin{equation*}
  j_0^{\rm
  h}(\mu,\gamma)>j_0^{\rm d}(\mu) \quad \wedge \quad j_0^{\rm
  hd}(\mu,\gamma)>j_0^{\rm d}(\mu). 
\end{equation*}
The limit case \tm{j_0^{\rm
  hd}(\mu,\gamma)=j_0^{\rm
  h}(\mu,\gamma)=j_0^{\rm d}(\mu) \equiv j_0^{\rm DH}(\mu)} \footnote{The
concurrence of three bifurcations is not a codimension-three point
since two fixed points (a homogeneous and an inhomogeneous steady state) are
involved.} can be
reformulated as a condition for the timescale of the inhibitor
$\gamma$ such that the condition can be tested for
different system sizes \cite{Bose.ea:00}:
\begin{equation}
  \label{eq:suffgamma}
  \gamma^{\rm hd}(i_0^{\rm d}(\mu))>\gamma^{\rm DH}(\mu).
\end{equation}
The above inequality  becomes clear, if one considers that the
oscillatory instability of the filament is shifted toward higher imposed
current density values when lowering $\gamma$, whereas the Hopf bifurcation
point of the homogeneous steady state behaves in the opposite way and
the spatial instability does not depend on $\gamma$.

\begin{figure}[!tbp]
  \begin{center}
    \epsfig{file=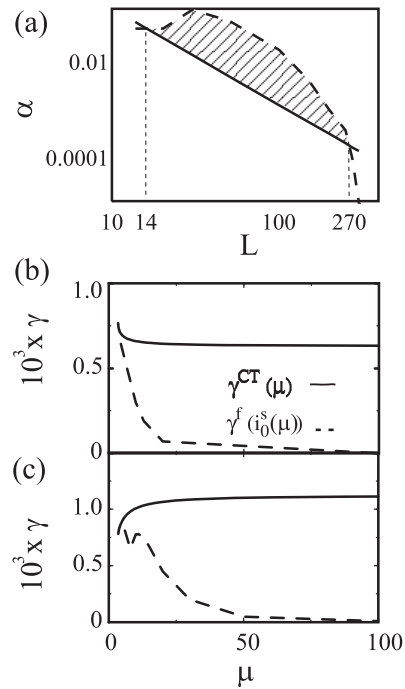,width=.29\textwidth}
    \caption{Thresholds for oscillatory instabilities at imposed
      current density values
      at which the homogeneous steady state becomes unstable with
      respect to spatial fluctuations (effective three parameter
      continuation) to test condition (\ref{eq:suffgamma}). The threshold
      for an oscillatory instability of the domain
      (\tm{\gamma^{\rm hd}(i_0^{\rm s}(\mu))}) and the
      codimension-two point (DH) analogous to a Turing-Hopf
      codimension-two bifurcation (\tm{\gamma^{\rm DH}(\mu)}) are
      shown as dashed and solid lines, respectively. In (a)
      the two curves are shown for the semiconductor system
      (double logarithmic plot) for 2d domains at the lower turning point of
      the current-voltage characteristic. The hatched region
      indicates the region in which the sufficient condition
      (\ref{eq:suffgamma}) for complex
      spatio-temporal dynamics is fulfilled. In (b) and (c)
      \tm{\gamma^{\rm hd}(i_0^{\rm s}(\mu))} and \tm{\gamma^{\rm
          DH}(\mu)} are shown for the electrochemical model for the
      two domain bifurcations at low and high current densities, respectively.}
    \label{fig:10}
  \end{center}
\end{figure}

In fig.\ \ref{fig:10}
both critical timescales are plotted for both
models. For the electrochemical model the critical timescales are also
shown for the upper part
of the S-shaped current-voltage characteristic. The above
arguments do equally apply for this region.
 As indicated by the hatched region for the semiconductor
system in fig.\ \ref{fig:10}a, condition (\ref{eq:suffgamma}) is fulfilled for a large interval of
system sizes $L$ (resp. \tm{\mu^{\rm (s)}}) for the lower part of the
S-shaped current-voltage characteristic.
Apart from spiking, a broad variety of periodic and chaotic
spatio-temporal modes has been found in this interval \cite{Bose.ea:00}.
Condition (\ref{eq:suffgamma}) is never found to hold
 for the upper part for the
semiconductor model (not shown). 
It can be seen in fig.\ \ref{fig:10}b and \ref{fig:10}c for the upper
and lower part of the S-shaped current-voltage characteristic, respectively, that 
condition (\ref{eq:suffgamma}) is  apparently never
fulfilled in the electrochemical system for any system size.

Thus also the absence of spiking in the electrochemical system is
easily understood; spiking evolves when the breathing mode eventually
comes very close to the plane of homogeneous modes which constitutes a
stable focus in this plane.
The relaxation close to 
the homogeneous fixed point in the plane of the homogeneous modes
leads to the small, almost homogeneous, oscillations and then the
spike evolves again as the trajectory leaves the plane of homogeneous
dynamics along the unstable direction of the homogeneous fixed point (cf.\ fig.\ \ref{fig:9}).
In the electrochemical system the plane of homogeneous modes always
constitutes an unstable focus for parameter values in which the domain
loses stability and thus the trajectory of inhomogeneous oscillations
never comes close to the unstable homogeneous fixed point.

\section{Conclusions}
The comparison of the two models presented in this paper allows us to
identify bifurcations that exist in bistable
systems subject to global inhibition. Apart
from electrochemical and semiconductor systems such dynamics might be
encountered in a variety of other systems, e.\ g.\ gas discharge
devices \cite{Raizer:97}.

Stationary large amplitude 
spatial patterns called domains or filaments appear via a subcritical
spatial bifurcation of the uniform state and form attractors in the whole range
of effective autocatalysis for common parameter values in such
systems. A characteristic length scale can
be defined that facilitates quantitative comparison of the respective
models.
For comparable timescales of activator and inhibitor
stable homogeneous relaxation oscillations can be expected.
For slow dynamics of the globally coupled inhibitor oscillatory
instabilities of the domains occur, initially near the turning points
of the current-voltage characteristic of the domain. However, the routes to
complex spatio-temporal patterns depend on the local dynamics and might
thus differ in each individual system under consideration.

We have identified the following scenarios: If the Hopf bifurcation of
the domain is supercritical, the system 
will display stable breathing domains. In the case of a subcritical
bifurcation the dynamics depends upon the further structure of the
bifurcation diagram. If condition (\ref{eq:suffgamma}) is fulfilled, the
onset of stable breathing or spiking modes can be expected. When inequality
(\ref{eq:suffgamma}) is not fulfilled and the oscillatory instability of
the domain is subcritical, no general statement regarding the
resulting dynamics is possible. Either homogeneous relaxation
oscillations or complex spatio-temporal dynamics may result in this
case.

We have demonstrated the above general statements with
two models exhibiting different scenarios leading to stable complex
spatio-temporal dynamics, thus illustrating the general
scheme. Condition (\ref{eq:suffgamma}), which ensures that stationary or
uniform modes are either unstable or 
do not exist, is fulfilled for the 
semiconductor system in a wide parameter range, but it can never be
satisfied in the specific electrochemical model
(\ref{eq:nondimphi}),(\ref{eq:nondimtheta}). As a consequence the  
electrochemical breathing current 
domains always coexist with homogeneous oscillations. Thus they have a
small basin of attraction compared to the 
situation in the semiconductor model (\ref{eq:sema}),(\ref{eq:semu}) where
no other mode is stable in a certain 
parameter range. As another consequence spiking current filaments are only present in the
semiconductor system.
Complex dynamics could
only be found near the turning point of the current-voltage
characteristic corresponding to the lower value
of the imposed current density in both systems.

These results emphasize the necessity to incorporate the spatial
degree of freedom when studying electrochemical systems with negative
differential resistance. Breathing current domains constitute a qualitatively new mode of
complex spatio-temporal dynamics in electrochemical systems reported
here for the first time. This
mode may evolve to chaotic spatio-temporal dynamics via a period
doubling cascade.

 It
should be noted that recent experimental studies  of the CO-oxidation on
Pt-single crystal electrodes have shown small amplitude oscillations
of the potential
in the range of negative differential resistance
\cite{Koper.ea:01}. This system might be an experimental illustration
of the above 
results, and therefore spatially resolved measurements
would be desirable.

\begin{acknowledgments}
We acknowledge financial support of the Deutsche
Forschungsgemeinschaft in the framework of the Sonderforschungsbereich
555 ``Complex Nonlinear Processes'', projects B4 and B1.
\end{acknowledgments}

\appendix

\section{Non-dimensionalization}
\label{app1}

In this section we give the transformations yielding the dimensionless
model equations (\ref{eq:functheta}),(\ref{eq:funcir}). In physical units
the equations read \cite{Mazouz.ea:00}
\begin{align*}
  {\rm C}_0 \frac{{\rm d} \pdl}{{\rm d} t} 
  &=
  {\rm i_0'} - {\rm \chi n F \bar{c}_{\rm r} k_{\rm r}} (1-<\theta>)
  \\
  & \quad
  \cdot \exp \left( \chi \frac{\rm \alpha n F}{ R T} ( \pdl -
    \widetilde{\rm V}) \right)
  \\
  \pdfdx{\theta}{t}
  &=
  {\rm k_{\rm ad} \bar{c}_{\rm ad}} (1-\theta) \exp \left(- \alpha
    w'(\pdl,\theta) \right) 
  \\
  & \quad
  - {\rm k_{\rm d}} \theta \exp \left( (1-\alpha)
    w'(\pdl,\theta) \right)
  \intertext{with}
  w'(\pdl,\theta)
  &=
  \frac{\rm C_0-C_1}{\rm 2
    N_{\rm max} k_{\rm B} T} \pdl^2  +      \frac{{\rm
    g'} \theta}{\rm R T}.
\end{align*}
The meaning of the numerous constants is given in
\cite{Mazouz.ea:00} and typical values of the constants are shown in 
table \ref{tab:1}. 
\begin{table}[!tbp]
\caption{Typical parameter values}
\label{tab:1}
\begin{ruledtabular}
  \begin{tabular}{r @{=} l c r @{=} l}
    $  {\rm k_{\rm ad}}    $ & $  1 \cdot 10^4 \frac{\rm cm^3}{\rm mol\ s}
    $ & \qquad \qquad& $
    {\rm C_0}           $ & $  20 \cdot 10^{-6} \frac{\rm C}{\rm V\ cm^2}
    $ \\ $
    {\rm k_{\rm d}}     $ & $  5 \cdot 10^{-3} \rm s^{-1}
    $ &  &$
    {\rm C_1}           $ & $  2 \cdot 10^{-6} \frac{\rm C}{\rm V\ cm^2}
    $ \\ $
    {\rm k_{\rm r}} \exp( - \frac{\rm \alpha n F}{\rm R T} \widetilde{\rm V})$ & $  2 \cdot 10^{-8}
    \frac{\rm cm}{\rm s}
    $ &  &$
    \bar{\rm c}_{\rm ad} $ & $  1 \cdot 10^{-6} \frac{\rm mol}{\rm cm^3}
    $ \\ $
    {\rm N_{\rm max}}   $ & $  1 \cdot 10^{14} {\rm cm^{-2}}
    $ &  &$
    {\rm D_{\theta}}    $ & $  1 \cdot 10^{-5} \frac{\rm cm^2}{\rm s}
    $ \\ $
    {\rm n}             $ & $  1
    $ &  &$
    {\rm g'}            $ & $  - 1.2 \cdot 10^5 \frac{\rm J}{\rm mol}
    $ \\ $
    \chi                $ & $  1
    $ &  &$
    \alpha              $ & $  1/2
    $ \\ $ 
    {\rm T}             $ & $  300 {\rm K}$
  \end{tabular}
\end{ruledtabular}
\end{table}
The model equations (\ref{eq:functheta}),(\ref{eq:funcir}) are 
retained via the transformations of the variables according to
\begin{align*}
  \pdl \rightarrow \pdl' &= \frac{\rm \alpha n F}{\rm R T} \pdl\\
  t \rightarrow t' &= \frac{\rm \pi^2 D_{\rm \theta}}{\rm L^2} t\\
  x  \rightarrow x' &= \frac{\pi}{\rm L} x,
\end{align*}
and with the introduction of the parameters
\begin{eqnarray*}
  \mu &=& \frac{\rm L^2 k_{\rm ad} \bar{\rm c}_{\rm ad}}{\pi^2 {\rm
      D}_{\rm \theta}}\\
  p   &=& \frac{\rm k_d}{\rm k_{\rm ad} \bar{\rm c}_{\rm ad}}\\
  \nu&=&\frac{\rm R^2 T (C_0-C_1)}{\rm 2 N_{\rm max}
    k_B \alpha n^2 F^2}\\
  g  &=& \frac{\alpha \rm g'}{\rm R T}\\
  \gamma & = & \frac{ \alpha (\rm L n F)^2}{\pi^2 \rm D_{\rm \theta} R T {\rm C}_0} \rm
  \bar{c}_{\rm r} k_{\rm r} \exp (- \chi \frac{\alpha n F}{R T}\widetilde{V})\\
  i_0 &=& \frac{\rm i_0'}{\rm n F \bar{c}_{\rm r} k_{\rm r} \exp 
    (- \chi \frac{\alpha n F}{R T} \widetilde{V})}.
\end{eqnarray*}

With the  values given in table \ref{tab:1} the parameters $p$, $\nu$ and
$g$ are fixed to
\mbox{$\nu=0.025$}~\footnote{Note that the value of $\nu$ was given as
  \mbox{$\nu=2$} in previous papers
  \cite{Mazouz.ea:00,Krischer.ea:00}, but the above value fits the 
  physical situation better.}
, \mbox{$p=0.5$} and \mbox{$g=-2.4$}. They correspond to such physical values
as free adsorption sites or interaction strength. 
$\gamma$ depends on the well accessible concentration of the reacting
species, which can be varied over several decades; \mbox{$\mu \simeq 100
  (L[cm])^2$}; $i_0$ can be set by the galvanostatic control unit and typical
values will be of order \mbox{$10^3$--$10^4$}.

\section{Bifurcations and phase portraits near the DH-codimension-two point}
\label{app2}
\begin{figure}[!tbp]
  \begin{center}
    \epsfig{file=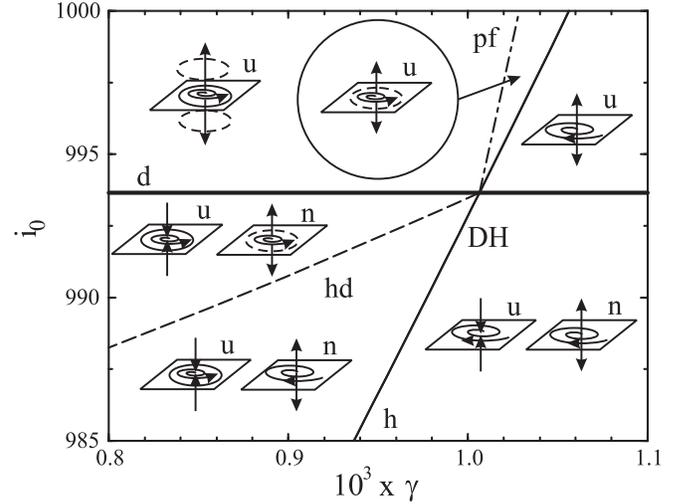,width=.48\textwidth}
    \caption{Bifurcations and projections of phase portraits close to
      the codimension-two domain-Hopf-bifurcation (DH). Thick solid line: 
      spatial instability of the homogeneous steady state (d). Thin
      solid line: Hopf bifurcation of the homogeneous steady state
      (h). Dashed line: Hopf bifurcation of the stationary
      unstable domain (hd). Dash-dotted line: pitchfork bifurcation of
      limit cycles that stabilizes the homogeneous limit cycle (pf). ``u''
      and ``n'' denote the planes of uniform and (nonuniform) domain modes, respectively.}
    \label{fig:11}
  \end{center}
\end{figure}

In fig.\ \ref{fig:11} 
the bifurcations and phase portraits near the codimension-two point in
which the domain bifurcation and Hopf bifurcation of the homogeneous steady
state meet (DH)
is shown. The additional branches not shown in fig.\ \ref{fig:8} are a
Hopf bifurcation of the unstable stationary domain leading to an
unstable inhomogeneous limit cycle and the pitchfork bifurcation of
periodic orbits that stabilizes the homogeneous limit cycle born in
the Hopf bifurcation of the homogeneous steady state and which is the
origin of another unstable inhomogeneous limit cycle
(cf.\ fig.\ \ref{fig:4}a). Both branches terminate in the DH. The respective
phase portraits (insets) depict the dynamics schematically in a projection
on the plane spanned by the eigenvectors of the two complex
conjugate eigenvalues describing the Hopf bifurcation of the homogeneous fixed point
resp.\ the stationary unstable domain. The third direction describes
the subcritical domain bifurcation (spatial mode).

\small

\end{document}